# A Study on the Point-Wise Time-Space Estimates of Fundamental Solution for Higher Order Schrödinger Equation Whose Symbol Function is One Kind of Degenerate Polynomial

*Kim Myong Jin,* Dr. *Kim Jin Myong*

College of Mathematics, **Kim Il Sung** University

**Abstract** In this paper we investigated the point-wise time-space estimates of the fundamental solution for higher order Schrödinger equation.
These estimates improve the result of Yao [2] and generalize the one of Cui [1].
**Key words** Schrödinger equation, point-wise time-space estimate

The great leader comrade **Kim Il Sung** said as follows.

"**We should further strengthen scientific research and rapidly develop science and technology so that any scientific and technical problems in economic construction can be solved as soon as they arise and thus successfully make the national economy scientific and fully ensure that it becomes Juche-orientated and modern.**"("**Kim Il Sung** Works" Vol. 35, 312p)

This paper is concerned with the estimates of fundamental solution for higher order Schrödinger equation.

In the case when symbol function is inhomogeneous elliptic non-degenerate polynomial, Cui [1] obtained the sharp point-wise time-space estimate of the fundamental solution for higher order Schrödinger equation. In 2008, Yao [2] introduced a kind of degenerate polynomial class and established a point-wise time-space estimate for those functions.

We improved the global estimate of Yao [2] to the best one and this estimate generalizes the result of Cui [1].

We consider the following equation.

$$\begin{cases} \dfrac{\partial u}{\partial t}(t,\, x) = iP(D)u(t,\, x) \\ u(0,\, x) = u_0(x) \end{cases},\ t \in \mathrm{R},\ x \in \mathrm{R}^n \qquad (1)$$

where $D = -i(\partial/\partial x_1, \cdots, \partial/\partial x_n)$, $P : \mathrm{R}^n \to \mathrm{R}$ is an inhomogeneous elliptic polynomial which satisfies the following condition.

Condition $(H_b)$ [2] $\{\lambda_k(\xi)\}_1^n$ are the eigen-values of Hessian of $P$ and $0 \le b \le 1$.

① $\max\limits_{1 \le k \le n} |\lambda_k(\xi)^{-1}| = O(|\xi|^{-(m-2)b})$  $(|\xi| \to \infty)$

② $\{\lambda_k(\xi)\}_1^n$ have the same sign.

When $b = 1$, $(H_1)$ is equivalent to the non-degenerate condition of $P$ [1].

In the rest of this paper we will denote by $C$ the generic constants independent of $t,\ x,\ \xi$.

By our assumptions, there exists a constant $L > 0$ such that $|\xi| \geq L$:
$$|\lambda_k(\xi)| \geq C|\xi|^{(m-2)b}, \quad |\nabla P(\xi)| \geq C|\xi|^{m-1}.$$

Using the Fourier transform, the solution of equation (1) is given by
$$u(t, x) = I(t, \cdot) * u_0(x), \quad u_0 \in S(\mathbf{R}^n), \quad I(t, x) := \int_{\mathbf{R}^n} e^{itP(\xi) + i\langle x, \xi \rangle} d\xi.$$

When $\Omega := \{\xi \in \mathbf{R}^n : |\xi| \geq L\}$,
$$I(t, x) = \int_{\Omega} e^{itP(\xi)+i\langle x, \xi\rangle} d\xi + \int_{\Omega^c} e^{itP(\xi)+i\langle x, \xi\rangle} d\xi = I_1(t, x) + I_2(t, x), \quad |I_2(t, x)| \leq C$$

is obvious.

**Theorem 1** Assume that $P$ is inhomogeneous elliptic polynomial and satisfies the condition $(H_b)$, $b \in [1/2, 1]$. Then there exist certain constants $L, C > 0$ such that

$$|I_1(t, x)| \leq \begin{cases} C|t|^{-\sigma}, & 0 < |t| < 1 \\ C|t|^{-n/2}, & |t| \geq 1 \end{cases} \quad (2)$$

where $\sigma = n/[(2b-1)(m-2) + 2]$.

**Proof** Write $\Omega = \bigcup_{j=1}^{3} \Omega_j$ where

$$\Omega_1 = \{\xi \in \Omega : |\xi| < |t|^{-1/m}\}, \quad \Omega_2 = \{\xi \in \Omega : |\xi| > |t|^{-1/m}/2, |\nabla P(\xi) + x/t| < |x/t|\},$$
$$\Omega_3 = \{\xi \in \Omega : |\xi| > |t|^{-1/m}/2, |\nabla P(\xi) + x/t| > |x/t|/2\}.$$

Choose the following partition of unity subordinate to this covering:
$$\varphi_1(\xi) = \varphi(\xi|t|^{1/m}), \quad \varphi_2(\xi) = (1-\varphi_1(\xi))\varphi((\nabla P(\xi)+x/t)/|x/t|), \quad \varphi_3(\xi) = 1 - \varphi_1(\xi) - \varphi_2(\xi)$$

where $\varphi \in C_0^\infty(\mathbf{R}^n)$ such that $\varphi(\xi) = \begin{cases} 1, & |\xi| < 1/2 \\ 0, & |\xi| > 1 \end{cases}$. When

$$I_1(t, x) = \sum_{j=1}^{3} I_{1j}(t, x), \quad I_{1j}(t, x) = \int_{\Omega} e^{itP(\xi)+i\langle x, \xi\rangle} \varphi_j(\xi) d\xi$$

One can easily obtain the following estimates.
$$|I_{11}(t, x)| \leq C|t|^{-n/m}, \quad |I_{13}(t, x)| \leq C|t|^{-n/m} \quad (3)$$

Let us estimate $I_{12}(t, x)$.

By the definition of $\Omega_2$, $|\nabla P(\xi)| \sim |\xi|^{m-1} \sim |x/t|$, $\forall \xi \in \Omega_2$ holds true, hence there exists certain constants $C_1, C_2 > 0$ such that $\Omega_2 \subset \{\xi \in \mathbf{R}^n : 2C_1 r < |\xi| < C_2 r\}$, $r := |x/t|^{\frac{1}{m-1}}$.

Assume that $L < C_1 r$ (If $L \geq C_1 r$, then it is easy to get $|I_{12}(t, x)| \leq C$). To this end, choose a finite set $\{\xi_v\} \subset S^{n-1}$ such that $|\xi_v - \xi_{v'}| \geq 1/4$ ($v \neq v'$), $\min|\xi - \xi_v| < 1/4$, $\xi \in S^{n-1}$.

Write $\Omega_2 = \bigcup_v \Omega_2^v$ where $\Omega_2^v := \{\xi \in \Omega_3 : |\xi/|\xi| - \xi_v| < 1/2\}$, and choose the following partition of unity subordinate to the covering: $\chi_v = \zeta_v / \sum \zeta_l$, $\zeta_v = \varphi(4|\xi/|\xi| - \xi_v|)$, $\xi \in \Omega_2$

Then $I_{12}(t, x)$ be divided into $I_{12} = \sum_v I_{12}^v$, $I_{12}^v(t, x) = \int_\Omega e^{itP(\xi)+i\langle x, \xi\rangle} \varphi_2(\xi)\chi_v(\xi) d\xi$.

To estimate $I_{12}^v$, we need the inequality

$$|\nabla P(\xi) - \nabla P(\xi')| \geq Cr^{b(m-2)} |\xi - \xi'|, \quad \xi, \xi' \in \Omega_3^v. \qquad (4)$$

If we consider a function $f_\eta(t) := <\eta, \nabla P(\xi' + t(\xi - \xi'))>$, $t \in [0, 1]$, then from the Newton-Leibniz theorem $f_h(1) - f_\eta(0) = \int_0^1 f'_\eta(t) dt = \int_0^1 <\eta, HP(\xi' + t(\xi - \xi'))(\xi - \xi')> dt$ holds and when

$$\eta = \frac{\xi - \xi'}{|\xi - \xi'|}, \quad |\nabla P(\xi) - \nabla P(\xi')| \geq |f_\eta(1) - f_\eta(0)| = |<\eta, HP(\bar{\xi}_\eta)>| \cdot |\xi - \xi'| = \left|\sum_{i=1}^n \lambda_i(\bar{\xi}_\eta) \eta_i^2\right| \cdot |\xi - \xi'|$$

also holds true. Since $\bar{\xi}_\eta$ belongs to convex hull of $\Omega_2^v$, by definition of $\Omega_2^v$, $|\bar{\xi}_\eta| \geq C_1 r > L$ holds and  by assumptions (4) holds true. Now let us take a point $\xi_0 \in \Omega_2^v$ such that $|\nabla P(\xi_0) + x/t| \leq Cr^{b(m-2)} |t|^{-\sigma/n}/4$ and then accordingly split the set $\Omega_2^v$ into two part

$$V_1 = \{\xi \in \Omega_2^v : |\xi - \xi_0| < |t|^{-\sigma/n}\}, \quad V_2 = \{\xi \in \Omega_2^v : |\xi - \xi_0| > |t|^{-\sigma/n}/2\}.$$

Then $I_{12}^v$ is divided into corresponding integral $I_{121}^v$ and $I_{122}^v$. One can easily verify that $|I_{121}^v| \leq C|t|^{-\sigma}$. write $V_2 = \bigcup_{j=1}^n W_j$, where $W_j = \left\{\xi \in V_2 : \left|\partial_j P(\xi) + \frac{x_j}{t}\right| > \frac{1}{\sqrt{2n}} \left|\nabla P(\xi) + \frac{x}{t}\right|\right\}$, and choose the following partition of unity subordinate to the covering

$$\eta_j = \theta_j \bigg/ \sum_{l=1}^n \theta_l, \quad j = \overline{1, n}, \quad \theta_j(\xi) = \psi\left(\frac{\sqrt{2n}(\partial_j P + x_j/t)}{|\nabla P(\xi) + x/t|}\right), \quad \xi \in \Omega_3$$

where $\psi \in C_0^\infty(\mathbb{R})$, $\psi(s) = \begin{cases} 1, & s \geq 1 \\ 0, & s \leq 1/2 \end{cases}$. Then $I_{122}^v$ is divided into $n$ part, and we will consider only one of them.

$$|I_{1221}^v| = \left|\int_\Omega e^{itP(\xi) + i<x, \xi>} D_*^n \phi \, d\xi\right|$$

where $\phi(\xi) = \varphi_2(\xi) \chi_v(\xi)[1 - \varphi(|t|^{\sigma/n}(\xi - \xi_0))] \eta_1(\xi)$ and $\mathrm{supp}\,\phi \subset W_1$.

By (4) and choose of $\xi_0$

$$|\nabla P(\xi) - \nabla P(\xi_0)| \geq Cr^{b(m-2)} |\xi - \xi_0| \geq Cr^{b(m-2)} |\xi - \xi_0|/2, \quad |\nabla P(\xi_0) + x/t| \leq Cr^{b(m-2)} |\xi - \xi_0|/4$$

holds and these two yields $|\nabla P(\xi) + x/t| \leq Cr^{b(m-2)} |\xi - \xi_0|/4$.

Define $h(\xi) = 1/g(\xi) = i(\partial_1 P(\xi) + x_1/t)$, then $|h(\xi)| \sim |\nabla P(\xi) + x/t| \geq Cr^{b(m-2)} |\xi - \xi_0|/4$ and

$$|D_*^n \phi| \leq C \sum_{k+j=n} |t|^{-n} |\xi|^{j(m-2)} |\partial_1^k \phi| \cdot |h^{-1}|^{n+j} \leq$$

$$\leq C \sum_{k+j=n} \sum_{i=0}^k |t|^{-n} |\xi|^{j(m-2)} |\xi|^{(m-2)(k-i)} |\xi - \xi_0|^{-i} |h^{-1}|^{2n-i} \leq \qquad (5)$$

$$\leq C|t|^{-n} |\xi - \xi_0|^{-2n} \sum_{i=0}^n |\xi|^{(m-2)[n(1-2b) - i + bi]}$$

where $b \geq 1/2$ ensures $n(1-2b) - i + bi \leq 0$, $i = \overline{1, n}$ and taking into account this, $|\xi| \sim |\xi_0|$ and $|\xi - \xi_0| \leq C|\xi|$ we can obtain the following estimate of $I_{12}$.

$$|I_{1221}^v| \leq C|t|^{-n}\left(\int_{|\xi-\xi_0|\geq |t|^{-\sigma/n}}|D_*^n\phi|\,d\xi\right) \leq C|t|^{-n}|t|^{\sigma[1+(2b-1)(m-2)]}\sum_{i=0}^{n}|t|^{\frac{\sigma(1-b)(m-2)i}{n}} \leq C|t|^{-\sigma}, \ 0<|t|<1$$

(6)

From (3), (6) the following estimate is obtained.

$$|I(t,x)| \leq C|t|^{-\sigma}, \ 0<|t|<1 \tag{7}$$

For the case of $|t|\geq 1$, $\Omega_1 \cap \Omega = \phi$ and hence $I_{11}(t,x)=0$.

Taking into account $|t|^{-1/m} \leq 1 \leq L$, $I_{13}(t,x)$ is estimated in the following.

$$|I_{131}| \leq C|t|^{-n}\int_{|\xi|>L}|\xi|^{-nm}\,d\xi \leq C|t|^{-n} \tag{8}$$

Take $\xi_0 \in \Omega_{12}^v$ such that $|\nabla P(\xi_0) + x/t| \leq Cr^{b(m-2)}|t|^{-1/2}/4$ and

$$V_1 = \{\xi \in \Omega_{12}^v : |\xi-\xi_0|<|t|^{-1/2}\}, \ V_2 = \{\xi \in \Omega_{12}^v : |\xi-\xi_0|>|t|^{-1/2}/2\},$$

then $|I_{121}^v| \leq C|t|^{-n/2}$ holds true. By definition of $\Omega_{12}$, $|\xi|>L$ holds and combining this with (5) one can obtain $|D_*^n\phi| \leq C|t|^{-n}|\xi-\xi_0|^{-2n}$, $|t|\geq 1$.

Hence

$$|I_{1221}^v| \leq C\int_{|\xi-\xi_0|>|t|^{-1/2}/2}|D_*^n\phi|\,d\xi \leq C\int_{|\xi-\xi_0|>|t|^{-1/2}/2}|t|^{-n}|\xi-\xi_0|^{-2n}\,d\xi \leq C|t|^{-n/2}, \ |t|\geq 1. \tag{9}$$

From (8), (9)

$$|I_1(t,x)| \leq C|t|^{-n/2}, \ |t|\geq 1. \tag{10}$$

Again from (7), (10) the proof is completed.□

**Theorem 2** Under the same assumption with theorem 1, there exists a certain constant $C>0$ such that $|I(t,x)| \leq \begin{cases} C|t|^{-\sigma}, & 0<|t|<1 \\ C, & |t|\geq 1 \end{cases}$ where $\sigma = n/[(2b-1)(m-2)+2]$.

**Remark** Yao [2] obtained the following estimate under $(H_b)$.

$$|I_1(t,x)| \leq \begin{cases} C|t|^{-\sigma_b}, & 0<|t|<1 \\ C|t|^{\rho_b}, & |t|\geq 1 \end{cases}, \ \rho_b = \frac{n[(m-3)-b(m-2)]}{(m-2)(2b-1)+2} \geq -\frac{n}{2}$$

Thus Theorem 1 improves (decreases) the superscript of time $t$ from $\rho$ in Yao [1] to the small value $-n/2$. In the meantime if we assume that $b=1$, the estimates is equal to the estimate of Cui [2] when $0<|t|<1$, and is better than $n/m$ ($m\geq 2$) of Cui [2] when $|t|\geq 1$. So our estimates become the best one which generalizes Cui [2].